\documentclass[notoc,11pt,paper]{JHEP3}

\date{March 2008}
\usepackage{epsfig}
\usepackage{latexsym}
\usepackage{amssymb}
\usepackage{booktabs}
\newcommand{\be}{\begin{equation}}
\newcommand{\ee}{\end{equation}}
\newcommand{\ba}{\begin{eqnarray}}
\newcommand{\ea}{\end{eqnarray}}
\newcommand{\bi}{\begin{itemize}}
\newcommand{\ei}{\end{itemize}}

\newcommand{\nn}{\nonumber \\}

\newcommand{\half}{{\textstyle\frac{1}{2}}}

\newcommand{\<}{\langle}
\renewcommand{\>}{\rangle}
\newcommand{\eq}{Eq.~}

\newcommand{\la}{\label}

\newcommand{\txts}{\textstyle}

\title{Energy-momentum tensor correlators and spectral functions}

\author{
Harvey~B.~Meyer\\
Center for Theoretical Physics\\ 
Massachusetts Institute of Technology\\
Cambridge, MA 02139, U.S.A.\\
E-mail: \email{meyerh@mit.edu}
}

\preprint{ MIT-CTP 3956}

\abstract{
We calculate the thermal Euclidean correlators and the
spectral functions of the energy-momentum tensor 
for pure gauge theories, including
at non-zero spatial momentum, at leading order in perturbation theory.
Our goal is to improve the extraction of transport properties
from Euclidean correlators that are computable in lattice QCD.
Based on our results and the predictions of hydrodynamics for the structure 
of the spectral functions at low frequencies, we show that 
the shear and bulk viscosities can advantageously be  extracted from the 
Euclidean correlators of the conserved charges, energy and momentum, 
at small but non-vanishing spatial momentum. The spectral functions in
these channels are free of the ultraviolet $\omega^4$
term which represents a large background 
to the thermal physics encoded in the correlators of the fluxes.
}

\begin{document}

\section{Introduction}
The particles produced in RHIC heavy-ion collisions
exhibit sizeable elliptic and radial 
flow~\cite{Arsene:2004fa,Back:2004je,Adcox:2004mh,Adams:2005dq}.
This is a signature of collective behavior, 
which the equations of ideal hydrodynamics have been able to 
describe quite successfully~\cite{Kolb:2000fha,Huovinen:2001cy,Teaney:2000cw}.
The leading corrections in a gradient expansion 
of flow velocities is parametrized by shear and bulk viscosity, $\eta$ and $\zeta$.
Detailed viscous relativistic hydrodynamics 
calculations~\cite{Luzum:2008cw,Romatschke:2007mq,Dusling:2007gi,Song:2007fn,Song:2007ux,Chaudhuri:2008sj}
find that the experimental data excludes 
a shear viscosity to entropy density ratio larger than about 0.3,
in agreement with early rough estimates~\cite{Teaney:2003kp}.
This would make the substance created an exceptionally  good fluid~\cite{Csernai:2006zz}, and
it is therefore of primary interest to improve our knowledge 
of the transport coefficients of the quark-gluon plasma.

Theoretically, the transport coefficients can be extracted
from the imaginary part of a retarded two-point correlation function; this imaginary part
is called the spectral function and is denoted by $\rho(\omega)$. This is the content
of the Kubo formulas~\cite{Kubo:1957mj}. By analytic continuation, the spectral function 
is also related to Euclidean correlators $C_{\rm E}$~\cite{Karsch:1986cq,Meyer:2008dq}, 
which are functions of Euclidean time $x_0$, spatial momentum ${\bf q}$ and 
temperature $T$, by
\be
C_{\rm E}(x_0,{\bf q}, T) = \int_0^\infty d\omega 
\rho(\omega,{\bf q}, T)\,
\frac{\cosh\omega(x_0-\frac{1}{2T})}{\sinh\frac{\omega}{2T}}.
\la{eq:C=intrho}
\ee
The spectral functions are odd in $\omega$ and have everywhere
the same sign as $\omega$.
Since viscosities parametrize the dissipation of momentum, 
the relevant operators are elements of the energy-momentum tensor $T_{\mu\nu}$.
Because the temperatures reached at RHIC are not much larger
than twice the critical temperature $T_c$ where a gas of hadrons
rapidly crosses over to a system with many more degrees of freedom 
(see~\cite{Karsch:2008fe} and Ref. therein), perturbative 
methods~\cite{Arnold:2006fz,Arnold:2000dr,Arnold:2003zc,Moore:2008ws,Aarts:2002cc} 
are not directly applicable. They can however
tell us about the asymptotic high $T$ behavior of the viscosities
and the associated spectral functions. They also allow us 
to familiarize ourselves with the intricate kinematics 
that arise at non-zero temperature and spatial momentum,
one of the objectives of this work. Finite-momentum meson spectral functions
were investigated analytically in~\cite{Aarts:2005hg}.

Lattice calculations of 
viscosities~\cite{Meyer:2007dy,Meyer:2007ic,Nakamura:2004sy,Karsch:1986cq}
have so far mostly focused on the Euclidean correlators 
of $T_{12}$ for the shear viscosity, and $T_{ii}$ (or $T_{\mu\mu}$)
for the bulk viscosity. Indeed the Kubo formulas read in these cases
\ba
\eta(T) &=& \pi\,\lim_{\omega\to 0} 
 \frac{\rho_{12,12}(\omega,{\bf 0},T)}{\omega}\,,
\la{eq:eta-old}
\\
\zeta(T) &=& \frac{\pi}{9}\lim_{\omega\to 0} 
 \frac{\rho_{ii,jj}(\omega,{\bf 0},T)}{\omega}\,,
\la{eq:zeta-old}
\ea
where $\rho_{\mu\nu,\rho\sigma}$ corresponds to
 $\<T_{\mu\nu}T_{\rho\sigma}\>$.
As we shall review in section 3, hydrodynamics, as an
effective theory describing low-frequency
phenomena around equilibrium, predicts the functional form 
of the spectral functions in these channels,
including at small but non-vanishing ${\bf q}$.

Leaving aside the problem of determining the spectral function 
given the Euclidean correlator, a particular difficulty 
in this approach is that the spectral functions grow 
as $\omega^4$ (times a power series in $\alpha_s$) at large
frequencies. This buries the contribution from the small 
$\omega$ region under a much larger contribution 
(by a factor of at least five) from ultraviolet modes.
The latter contribution is almost temperature independent and therefore 
does not advance in any way our understanding of thermal physics.
Given that the Euclidean correlators are determined by Monte-Carlo
methods and carry statistical errors, this is a significant drawback.
The difficulty is far more severe than in studies of the 
charmonium spectral functions~\cite{Asakawa:2003re,Aarts:2006nn,Aarts:2007pk}, 
or calculations of the 
electromagnetic conductivity~\cite{Aarts:2007wj}, because the spectral function
only grows as $\omega^2$ for the vector current $\bar\psi\gamma_i\psi$.

Methods to subtract the ultraviolet contribution
to the Euclidean correlator, and hence to enhance the 
sensitivity of the lattice observables to the low-frequency region described by 
hydrodynamics, have been proposed
and implemented~\cite{Meyer:2008dq}. One of these methods 
(subtracting the $T=0$ spectral function)
has the virtue of removing the $\omega^4$ contribution
completely, by contrast with a perturbative order-by-order
subtraction. The drawback is that positivity of the integrand
in \eq\ref{eq:C=intrho} is given up, and a large part of the 
signal is lost in the difference, an unfavorable situation
from the numerical point of view.

Here, based on the exact Ward identities that follow
from the conservation of the energy-momentum tensor,
we show that the spectral function of the energy density operator
with non-vanishing spatial momentum
goes to a constant as $\omega\to\infty$. 
Similarly,
the two-point function of the momentum density operator
grows only as $\omega^2$. 
This is confirmed by our  perturbative calculation.
These Euclidean correlators 
are therefore far more sensitive to the thermal effects
than the correlators of the fluxes.
Yet provided ${\bf q}$ is sufficiently small,
their low-frequency region is still described by hydrodynamics,
and therefore the shear and bulk viscosity can 
be extracted from them. 

We collect leading perturbative results on the Euclidean 
correlators and spectral functions in section 2, and discuss
in section 3 the interplay of the perturbative predictions with 
the hydrodynamics predictions to propose a new way 
to extract shear and bulk viscosity from 
Euclidean correlators.

\section{Perturbative calculation}
The Euclidean energy-momentum tensor for SU($N_c$) gauge theories reads
\ba
T_{\mu\nu}(x) &=& \theta_{\mu\nu}(x) + \frac{1}{4}\delta_{\mu\nu}\,\theta(x)
\\
\theta_{\mu\nu}(x) &=&
 {\txts\frac{1}{4}}\delta_{\mu\nu}F_{\rho\sigma}^a F_{\rho\sigma}^a
   - F_{\mu\alpha}^a F_{\nu\alpha}^a
\\
\theta(x) &=& {\txts\frac{\beta(g)}{2g}} ~ F_{\rho\sigma}^a  F_{\rho\sigma}^a
\\
\beta(g) &=&  -b_0g^3+\dots,\qquad  b_0={\txts\frac{11N_c}{3(4\pi)^{2}}}\,.
\ea
In the U(1) case, the summation over the adjoint index $a$ is trivial
and of course $\beta(g)=0$. In contrast with Minkovsky space,
 $T_{0i}=\theta_{0i}$ is an antihermitian operator. 
In particular, $\<T_{0i}(x)T_{0i}(0)\><0$
for $x\neq0$, and $P_j = i\int d^3{\bf x}\, T_{0j}(x) $ 
is the usual momentum operator, for instance $P_j|{\bf q}\> = q_j|{\bf q}\>$
for a one-particle state.

In view of the form of the energy-momentum tensor, we consider now 
the connected correlators of the field strength
tensor and those of the field strength tensor squared, 
at leading order in perturbation theory. The expectation
values obtained in that approximation are denoted by $\<\dots\>_0$
and throughout this paper we keep only connected diagrams.
The effect on spectral functions
of using Hard-Thermal Loop resummed 
perturbation theory~\cite{Braaten:1989mz}
in the region $\omega<T$ is beyond the scope of this work.
Nevertheless our results are perfectly physical, 
since they are exact in the U(1) case.

In  Feynman gauge, we have 
\ba
\<  F_{\mu\nu}(x) F_{\alpha\beta}(y)\>_0
&=& d_A\,\phi_{\mu\nu\alpha\beta}(x-y)\,, \\
 \phi_{\mu\nu\alpha\beta}(x) &\equiv &
   \delta_{\nu\beta}  f_{\mu\alpha}(x)
+ \,\delta_{\mu\alpha} f_{\nu\beta}(x)
- \,\delta_{\mu\beta} f_{\nu\alpha}(x)
- \,\delta_{\nu\alpha} f_{\mu\beta}(x)\,, \\
 f_{\mu\alpha}(x) &\equiv &
T\sum_{p_0} \int \frac{d^3{\bf p}}{(2\pi)^3} 
\frac{e^{i(p_0x_0+{\bf p\cdot x})}}{p_0^2+{\bf p}^2}\,p_\alpha \, p_\mu\,.
\la{eq:basic}
\ea
One then easily obtains
\be
\<   F^a_{\mu\nu}(x)  F^a_{\rho\sigma}(x)~
     F^b_{\alpha\beta}(y)  F^b_{\gamma\delta}(y) \>_0
 = d_A\Big[ \phi_{\mu\nu\alpha\beta}(x-y) \phi_{\rho\sigma\gamma\delta}(x-y)
  +   \phi_{\mu\nu\gamma\delta}(x-y)\phi_{\rho\sigma\alpha\beta}(x-y)\Big]\,.
\ee
The color factor is $d_A\equiv N_c^2-1$ for SU($N_c$) and 1 for U(1).
To study mixed correlators (which are functions of $(x_0,{\bf p})$),
we introduce the spatial Fourier transform of $\phi_{\mu\nu\alpha\beta}(x)$,
$\tilde \phi_{\mu\nu\alpha\beta}({\bf p},x_0) =
\int d^3{\bf x}\,\phi_{\mu\nu\alpha\beta}(x) ~e^{i{\bf p\cdot x}}$.
Then
\ba
&& \int d^3{\bf y}\, e^{i{\bf q}\cdot{\bf y}}~
\<\,  F^a_{\mu\nu}(0) F^a_{\rho\sigma}(0)~
 F^b_{\alpha\beta}(x_0,{\bf y}) F^b_{\gamma\delta}(x_0,{\bf y}) \,\>_0 \\
&& = d_A\int \frac{d^3{\bf p}}{(2\pi)^3}
\Big[\tilde\phi_{\mu\nu\alpha\beta}({\bf p},x_0)
 \tilde\phi_{\rho\sigma\gamma\delta}(-({\bf p+q}),x_0)
   + \tilde \phi_{\mu\nu\gamma\delta}({\bf p},x_0)\tilde
 \phi_{\rho\sigma\alpha\beta}(-({\bf p+q}),x_0)\Big].
\nonumber
\ea
The integral is worked out in appendix~\ref{app:CE}. The results are of 
course gauge-invariant. In the next
two subsections, we present the results in a number of 
channels of physical interest. Throughout this section, 
we align the spatial momentum with the $z$-axis, 
${\bf q}=q\hat e_3$, $q\geq 0$. Since the spectral function 
is odd in $\omega$, we also choose $\omega\geq0$
without loss of generality. The most general results
are given at the end of the section, however we find it useful
to give the explicit form in the simpler cases of zero temperature
or zero momentum.
\subsection{Zero temperature}
At $T=0$ and zero spatial momentum, 
by dimensional analysis, the correlation functions at tree-level 
all fall off as $1/x_0^5$. With finite-momentum and for $x_0>0$,
\ba 
\int d^3{\bf x}\, e^{i{\bf q\cdot x}} \<T_{13}(x)T_{13}(0)\>_0
&=& \frac{d_Ae^{-qx_0}}{5(4\pi)^2 x_0^5}\,(qx_0+2)\,(q^2x_0^2+3qx_0+6)\,,~
\\ 
\rho_{13,13}(\omega,q) &=& \frac{d_A\theta(\omega-q)}{10(4\pi)^2}\,\omega^2(\omega^2-q^2)\,,
\\ 
\int d^3{\bf x}\, e^{i{\bf q\cdot x}} \<T_{12}(x)T_{12}(0)\>_0
&=& \frac{4d_Ae^{-qx_0}}{5(4\pi)^2 x_0^5}\,(3+qx_0+q^2x_0^2)
\\ 
\rho_{12,12}(\omega,q) &=& \frac{d_A\theta(\omega-q)}{10(4\pi)^2}(\omega^2-q^2)^2\,,
\\ 
\int d^3{\bf x}\, e^{i{\bf q\cdot x}} \<\theta_{33}(x)\theta_{33}(0)\>_0
&=&  \frac{2d_Ae^{-qx_0}}{15(4\pi)^2 x_0^5}\,
(24+24qx_0+12q^2x_0^2+4q^3x_0^3+q^4x_0^4),~
\\ 
\rho_{33,33}(\omega,q) &=& 
 \frac{2d_A\theta(\omega-q)}{15(4\pi)^2}   \omega^4,
\\ 
\int d^3{\bf x}\, e^{i{\bf q\cdot x}} \<\theta(x)\theta(0)\>_0
&=& \left(\frac{11\alpha_sN_c}{6\pi}\right)^2
\frac{2d_Ae^{-qx_0}}{(4\pi)^2 x_0^5} \,(3+3qx_0+q^2x_0^2)\,,
\\ 
\rho_{\theta,\theta}(\omega,q) &=& \left(\frac{11\alpha_sN_c}{6\pi}\right)^2
\frac{d_A\theta(\omega-q)}{4(4\pi)^2}(\omega^2-q^2)^2.
\\ 
\int d^3{\bf x}\, e^{i{\bf q\cdot x}} \<\theta_{00}(x)\theta_{00}(0)\>_0
&=& 
\frac{2d_Aq^4e^{-qx_0}}{15(4\pi)^2 x_0} 
\\ 
\rho_{00,00}(\omega,q) &=& 
\frac{2d_A\theta(\omega-q)}{15(4\pi)^2}\,q^4.
\ea 
\subsection{Finite temperature}
The ${\bf q}=0$ correlation functions of interest are
\ba
\int d^3{\bf x} ~\<\, \theta_{00}(0)\,  \theta_{00}(x) \, \>_0 
&=&  \frac{4\pi^2d_AT^5}{15}\,,   \\
\int d^3{\bf x}~ \<\, \theta(0) \, \theta_{00}(x)\,  \>_0   & = &0\,, \\
\int d^3{\bf x}~ \<\, \theta(0) \,  \theta(x) \, \>_0  & = & 
\left(\frac{11\alpha_sN_c}{6\pi}\right)^2
 \frac{16d_AT^5}{\pi^2}\Big(f(\tau)-\frac{\pi^4}{60} \Big)\,,  \la{eq:SS} \\
\int d^3{\bf x} ~\<\, \theta_{12}(0)\, \theta_{12}\, \>_0 
& = & \frac{32d_AT^5}{5\pi^2} ~ \Big(f(\tau)-\frac{\pi^4}{72} \Big)\,,
 \la{eq:12;12} 
 \\  
 \int d^3{\bf x} ~\<\, \theta_{11}(0)\, \theta_{11}\, \>_0 
 &  =  & \frac{128d_AT^5}{15\pi^2} ~ \Big(f(\tau)-\frac{\pi^4}{96} \Big)\,,
\ea
where $\tau=1-2Tx_0$ and
\be
f(\tau) = \int_0^\infty ds~  s^4 ~ \frac{\cosh^2\tau s}{\sinh^2 s} =
\frac{\pi^4}{60} +
 \frac{3}{4}\sum_{n\geq 1}
  \frac{n}{(n-\tau)^5} + \frac{n}{(n+\tau)^5}.
\ee
The two point function of $\int d^3{\bf x}\,T_{00}(x)$ is 
time-independent, as expected for a conserved charge.
The spectral functions in the tensor and scalar channels read
(see for instance~\cite{Meyer:2007fc,Meyer:2007dy})
\ba
\rho_{12,12}(\omega,{\bf 0},T)&=& \frac{d_A}{10(4\pi)^2}\frac{\omega^4}{\tanh\frac{\omega}{4T}}
+\left(\frac{2\pi}{15}\right)^2 d_A T^4\,\omega \delta(\omega)
\la{eq:1212q0}
\\
\rho_{\theta,\theta}(\omega,{\bf 0},T) &=&
\frac{d_A}{4(4\pi)^2}  \left(\frac{11\alpha_sN_c}{6\pi}\right)^2
\frac{\omega^4 }{\tanh\frac{\omega}{4T}}.
\la{eq:ththq0}
\ea
\subsubsection{Finite momentum}
For a polynomial $P$, we define
\be
{\cal I}([P],\omega,q,T) =
\theta(\omega-q) \int_0^1  dz \frac{P(z)\,
\sinh\frac{\omega}{2T}}{\cosh\frac{\omega}{2T}-\cosh \frac{q z}{2T}}
+ \theta(q-\omega) \int_1^\infty dz \frac{P(z)\,
\sinh\frac{\omega}{2T}}{\cosh\frac{qz}{2T}-\cosh \frac{\omega}{2T}}.
\ee
Then the spectral functions read
\ba 
\rho_{13,13}(\omega,q,T) &=& \frac{d_A   }{8\,(4\pi)^2}
~ \omega^2(\omega^2-q^2)~~  {\cal I}([1-z^4], \omega, q, T)\,,
\la{eq:sfqfirst}
\\ 
\rho_{12,12}(\omega,q,T) &=& \frac{d_A}{32\,(4\pi)^2} ~ (\omega^2-q^2)^2
~~ {\cal I}([1+6z^2+z^4], \omega, q, T)\,,
\\ 
\rho_{33,33}(\omega,q,T) &=& \frac{d_A}{4\,(4\pi)^2}
\omega^4 ~~ {\cal I}([(1-z^2)^2], \omega, q, T)\,,  
\\ 
\rho_{\theta,\theta}(\omega,q,T) &=&  \!\!
\left(\frac{11\alpha_sN_c}{6\pi}\right)^2
\frac{d_A }{4\,(4\pi)^2}~(\omega^2-q^2)^2 ~~  {\cal I}([1], \omega, q, T)\,,
\\ 
\rho_{00,00}(\omega,q,T) &=&  
\frac{d_A }{4\,(4\pi)^2}~q^4 ~~  {\cal I}([(1-z^2)^2], \omega, q, T)\,,
\la{eq:00,00e3}
\\ 
-\rho_{01,01}(\omega,q,T) &=&  
\frac{d_A }{8\,(4\pi)^2}~q^2(\omega^2-q^2) ~~  {\cal I}([(1-z^4)], \omega, q, T)\,,
\la{eq:01,01e3}
\\ 
-\rho_{03,03}(\omega,q,T) &=&  
\frac{d_A }{4\,(4\pi)^2}~q^2\omega^2 ~~  {\cal I}([(1-z^2)^2], \omega, q, T)\,.
\la{eq:sfqlast}
\ea 
One finds that
\ba
 {\cal I}([1], \omega, q, T) =
-\frac{\omega}{q} \theta(q-\omega)
\,+\, \frac{2T}{q}\,\log\frac{\sinh(\omega+q)/4T}{\sinh{|\omega-q|/4T}}\,
\la{eq:sfelem}
\ea
and hence the trace anomaly correlator can be expressed in terms of elementary functions.
The integrals with $z^2$ and $z^4$ in the numerator can be expressed in terms of polylogarithms,
explicit formulas are given in appendix~\ref{app:SF}. A few spectral functions are displayed on Figure 1.

\FIGURE[t]{
\centerline{\includegraphics[width=8.4 cm,angle=-90]{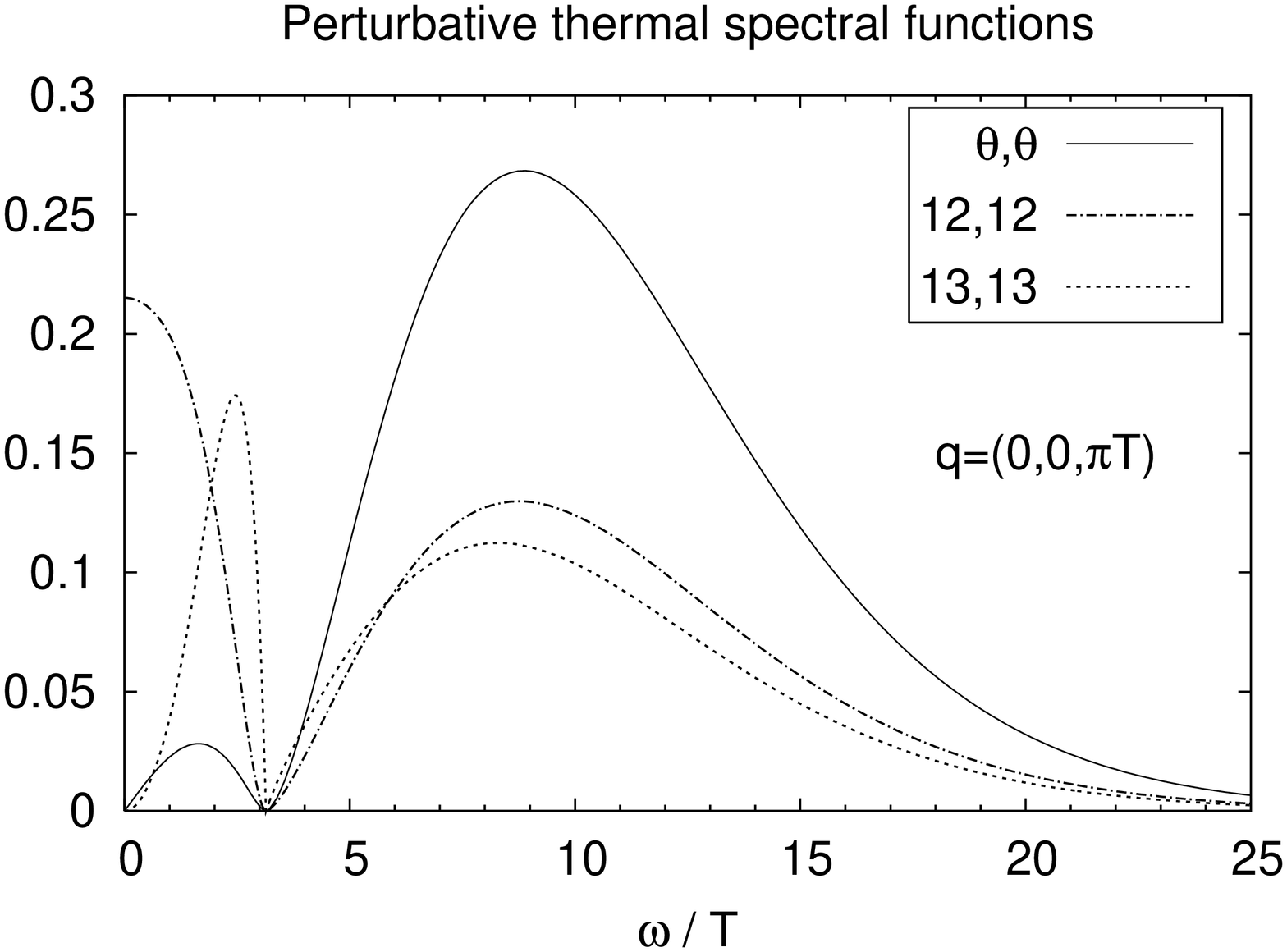}}
\caption{The function 
$\frac{1}{d_AT^4}[\rho(\omega,{\bf q},T)/\tanh(\omega/2T)-\rho(\omega,{\bf q},0)]$
for $q=\pi T$. For the trace anomaly $\theta$, 
the factor $(11\alpha_sN_c/6\pi)^2$ has been dropped.}
\label{fig:sfq}
}
\noindent

\section{Physics discussion}
At  small momentum and frequency,
the expression for the spectral functions of the momentum
densities are predicted by hydrodynamics 
(see~\cite{Teaney:2006nc} for an explicit derivation),
\ba
-\frac{\rho_{01,01}(\omega,{\bf q})}{\omega} 
&~\stackrel{\omega,q\to 0}{\sim} ~&
\frac{\eta}{\pi} \frac{q^2}{\omega^2+(\eta q^2/(Ts))^2}\,,
\la{eq:12hydro}   \\
-\frac{\rho_{03,03}(\omega,{\bf q})}{\omega} 
& ~\stackrel{\omega,q\to 0}{\sim}~  & \frac{\frac{4}{3}\eta+\zeta}{\pi}
\frac{\omega^2q^2}{(\omega^2-v_s^2q^2)^2+(\omega q^2 (\frac{4}{3}\eta+\zeta)/(Ts))^2}\,,
\la{eq:11hydro}
\ea
where $s$ is the entropy density, $v_s$ is the velocity of sound
and ${\bf q}=q\hat e_3$.

Based on the fact the matrix elements of $\partial_\mu T^{\mu\nu}$ vanish
between any two on-shell states, the Euclidean  correlators of the charges
and those of the fluxes are related in a simple way.
We emphasize that these relations are exact, since they derive from a Ward identity.
In terms of the spectral functions they read (${\bf q}=q\hat e_3$)
\ba
\omega^4\,\rho_{{00},{00}}(\omega,{\bf q}) & =& q^4 \rho_{33,33}(\omega,{\bf q})
\\
-\omega^2\,\rho_{{01},{01}}(\omega,{\bf q}) & =& q^2\,\rho_{13,13}(\omega,{\bf q})
\la{eq:0202e1}
\\
-\omega^2\,\rho_{{03},{03}}(\omega,{\bf q}) & =& q^2\,\rho_{33,33}(\omega,{\bf q}).
\la{eq:0101e1}
\ea
These relations are in particular satisfied by our treelevel expressions, 
\eq(\ref{eq:sfqfirst}--\ref{eq:sfqlast}). Note that the minus signs on the right-hand side
of  \eq(\ref{eq:0202e1}--\ref{eq:0101e1}) are absent 
in Minkovsky space (they come from the definition of $T_{0k}$ itself, 
see the remark at the beginning of section 2).

Equations (\ref{eq:0202e1}--\ref{eq:0101e1}) and (\ref{eq:12hydro}--\ref{eq:11hydro})
can be combined to obtain (\ref{eq:eta-old}) and (\ref{eq:zeta-old}),
which have so far been the basis of the calculation of shear and bulk viscosity
using lattice Monte-Carlo techniques~\cite{Meyer:2007dy,Meyer:2007ic,Nakamura:2004sy,Karsch:1986cq}. 
In that strategy, the momentum
${\bf q}$ is set to zero at the outset, and $\omega$ is sent to zero at the end.

However, in view of \eq(\ref{eq:12hydro}--\ref{eq:11hydro}),
the shear and bulk viscosity can be extracted from the low-freqency behavior
of the spectral functions for the four charge densities $T_{0\mu}$, as long as ${\bf q}\neq0$.
The advantage of using these correlators is that the ultraviolet contributions 
are highly suppressed compared to the correlators of the spatial components. 
We may illustrate this point by two numerical examples.

In \eq\ref{eq:1212q0}, relevant to shear viscosity,
the $\omega^4/\tanh\frac{\omega}{4T}$ term  contributes 
for 86\% to the Euclidean correlator at $t=1/2T$.
By contrast, in the $\rho_{01,01}$ channel for $q=\pi T/2$, 
the contributions to  $C_{\rm E}(x_0=1/2T)/(d_AT^5)$ 
coming from $\omega>q$ and $\omega<q$ are respectively 
$\approx0.04$ and 0.6, assuming the treelevel form 
(\eq\ref{eq:01,01e3}) in the first region
and \eq\ref{eq:12hydro} with $s=\frac{3}{4}s_{\rm SB}=\frac{1}{15}d_A\pi^2T^3$
and $\eta/s=1/4\pi$ in the second.

In the energy density channel, the increase in sensitivity to the low-frequency 
region  of the spectral function is even more dramatic: 
for  $q=\pi T/2$, the contribution to $C_{\rm E}(x_0=1/2T)/(d_AT^5)$ 
from $\omega>q$ is merely $\approx0.01$ based on \eq\ref{eq:00,00e3}, 
while the sub-threshold contribution is about $1.9$
for $s=\frac{3}{4}s_{\rm SB}$, 
$v_s^2=\frac{1}{3}$, $\eta/s=1/4\pi$ and $\zeta=0$.
These values are inspired by the strongly coupled
 ${\cal N}=4$ SU($N_c$) SYM gauge theory, 
which can be studied by analytic AdS/CFT methods
 (see for instance~\cite{Son:2007vk} and Ref. therein).

One of the key issues in practice is to achieve sufficiently small  $q$ 
for the hydrodynamics prediction to be valid below threshold.
One will want to reach $q<\pi T/2$ and check this explicitly.
Since $q=2\pi /L$ is the smallest momentum available in a periodic box, 
this requires simulating in rather large spatial volumes.
Approaching a second order phase transition, 
it may become impractical to reach sufficiently small momenta.
Anisotropic lattices~\cite{Sakai:2007cm, Meyer:2008dq} 
can help to achieve large physical volumes
while keeping discretization errors under control.
From the algorithmic point of view, having a non-zero spatial momentum
is a particularly favorable situation for the 
multi-level algorithm~\cite{Meyer:2002cd,Meyer:2003hy,Majumdar:2003xm}, 
since it allows the ultraviolet fluctuations to be tamed very efficiently.
Indeed non-perturbative, non-zero momentum correlators 
of the momentum fluxes were presented with 
1\% precision in~\cite{Meyer:2008dq}.

In the case of the vector current $\bar\psi\gamma_\mu\psi$,
it may also be profitable to exploit the correlator
of the charge density with non-zero momentum.
Non-zero momentum correlators of the current
have been computed with good precision~\cite{Aarts:2006wt},
exploiting twisted boundary conditions  
to scan low momenta more easily.

\acknowledgments
This work was supported in part by
funds provided by the U.S. Department of Energy under cooperative research agreement
DE-FG02-94ER40818.
\appendix
\section{Calculation of Euclidean correlators\la{app:CE}}
In this appendix we  derive a form  of the Euclidean correlators
which is well suited for accurate numerical evaluation.
We start with $\<T_{12}(0)\,T_{12}(x)\>  $, and then give
results for other channels treated in the same way.
The orientation of the momentum is specified by 
${\bf q}=q\hat e_i$, $i=1,2,3$. The color factor $d_A$ is omitted.

\subsection{The $\<T_{12}(0)\,T_{12}(x)\>  $ correlator}
\ba
\<T_{12}(0)\,T_{12}(x)\>
&=& \<(F_{10}F_{20})(0)\,(F_{10}F_{20})(x)\> 
+ \<(F_{13}F_{23})(0)\,(F_{13}F_{23})(x)\> 
\nn
&& +2\, \<(F_{10}F_{20})(0)\,(F_{13}F_{23})(x)\> .
\ea
Explicitly, the tensor $\tilde \phi_{\mu\nu\alpha\beta}$ reads
\ba
\tilde \phi_{\mu\nu\alpha\beta}({\bf p},x_0) &= &
 \frac{1}{L_0}\sum_{p_0}
  \frac{e^{ip_0x_0}}{ p_0^2+{\bf p}^2} 
 \Big[\delta_{\nu\beta}  \,  p_\mu \,  p_\alpha
     + \,\delta_{\mu\alpha}  \,  p_\nu  \,  p_\beta 
-  \delta_{\mu\beta} \,   p_\nu \,  p_\alpha
     - \,\delta_{\nu\alpha} \,  p_\mu \,  p_\beta \Big].
\nonumber
\ea
In the $T\to0$ limit of course one can make 
the substitution $\frac{1}{L_0}\sum_{p_0}\to\int\frac{dp_0}{2\pi}$.
Defining
\ba
{\cal A}&=& \frac{1}{L_0^2}\sum_{p_0,q_0}\int\frac{d^3{\bf p}}{(2\pi)^3}
\, (p_1^2+p_0^2) \frac{e^{ip_0x_0}}{p_0^2+{\bf p}^2}
\,((p_2+q_2)^2+q_0^2)\frac{e^{iq_0x_0}}{q_0^2+({\bf p+q})^2}\,,
\\
{\cal B}&=& \frac{1}{L_0^2}\sum_{p_0,q_0}\int\frac{d^3{\bf p}}{(2\pi)^3}
\, p_1 p_2 \frac{e^{ip_0x_0}}{p_0^2+{\bf p}^2}
\,(p_1+q_1)(p_2+q_2)\frac{e^{iq_0x_0}}{q_0^2+({\bf p+q})^2}\,,
\\
{\cal C}&=& \frac{1}{L_0^2}\sum_{p_0,q_0}\int\frac{d^3{\bf p}}{(2\pi)^3}
\, (p_1^2+ p_3^2) \frac{e^{ip_0x_0}}{p_0^2+{\bf p}^2}
\,[(p_2+q_2)^2+(p_3+q_3)^2]\frac{e^{iq_0x_0}}{q_0^2+({\bf p+q})^2}\,,
\\
{\cal D}&=& \frac{-1}{L_0^2}\sum_{p_0,q_0}\int\frac{d^3{\bf p}}{(2\pi)^3}
\, p_0 p_3 \frac{e^{ip_0x_0}}{p_0^2+{\bf p}^2}
\,q_0(p_3+q_3)\frac{e^{iq_0x_0}}{q_0^2+({\bf p+q})^2}\,,
\ea
we have 
\ba
\int d^3{\bf x}\,e^{i{\bf q\cdot x}}
\< F_{10}F_{20})(0)\,(F_{10}F_{20})(x)\>
&=& {\cal A} + {\cal B}\,,
\\
\int d^3{\bf x}\,e^{i{\bf q\cdot x}}
\< F_{13}F_{23})(0)\,(F_{13}F_{23})(x)\>
&=& {\cal B} + {\cal C}\,,
\\
\int d^3{\bf x}\,e^{i{\bf q\cdot x}}
\< F_{10}F_{20})(0)\,(F_{13}F_{23})(x)\> &=& {\cal D}\,.
\ea
Using the Poisson summation formula, we obtain
\ba
{\cal A} &=& \frac{1}{4}\sum_{m,n\in {\bf Z}} 
\int\frac{d^3{\bf p}}{(2\pi)^3}\,
\frac{(p_1^2-{\bf p}^2) \, ((p_2+q_2)^2-({\bf p+q})^2)}{|{\bf p}|\,|{\bf p+q}|}
\,e^{-|{\bf p}||x_0+mL_0|} \, e^{-|{\bf p+q}||x_0+nL_0|}\,,
\\
{\cal B} &=& \frac{1}{4}\sum_{m,n\in {\bf Z}}
\int\frac{d^3{\bf p}}{(2\pi)^3}\,
\frac{p_1p_2\,(p_1+q_1)(p_2+q_2)}{|{\bf p}|\,|{\bf p+q}|}
\,e^{-|{\bf p}||x_0+mL_0|} \, e^{-|{\bf p+q}||x_0+nL_0|}\,,
\\
{\cal C} &=& \frac{1}{4}\sum_{m,n\in {\bf Z}}
\int\frac{d^3{\bf p}}{(2\pi)^3}\,
\frac{(p_1^2+p_3^2)\,[(p_2+q_2)^2+(p_3+q_3)^2]}{|{\bf p}|\,|{\bf p+q}|}
\,e^{-|{\bf p}||x_0+mL_0|} \, e^{-|{\bf p+q}||x_0+nL_0|}\,,
\\
{\cal D} &=& \frac{1}{4}\sum_{m,n\in {\bf Z}} 
s_m({\txts\frac{x_0}{L_0}})\,s_n({\txts\frac{x_0}{L_0}})
\int\frac{d^3{\bf p}}{(2\pi)^3}\, p_3(p_3+q_3)\, 
\,e^{-|{\bf p}||x_0+mL_0|} \, e^{-|{\bf p+q}|\,|x_0+nL_0|}\,,
\la{eq:poisson}
\ea
where 
\be
s_n(x) \equiv {\rm sign}(x+n)\,.
\ee
Integrating in spherical coordinates leads to
\ba
\int d^3x\,e^{i{\bf q\cdot x}}\<T_{12}(0)\,T_{12}(x)\> 
=
\frac{q^5}{64\pi^2}
\sum_{m,n\in{\bf Z}}
\int_0^\infty k^3 dk\,e^{-|x_0+mL_0|k}~\times
\nonumber
\\
\times~ \Big\{ k\sqrt{k^2+1}\,
( J_{\cal A} +2\,J_{\cal B} +J_{\cal C})
\left(q|x_0+nL_0|\sqrt{k^2+1},\,\frac{2k}{k^2+1},k\right) 
\\
+2(k^2+1)\,
s_m({\txts\frac{x_0}{L_0}})\,s_n({\txts\frac{x_0}{L_0}})
\, I_{\cal D}\left(q|x_0+nL_0|\sqrt{k^2+1},\frac{2k}{k^2+1},k\right)\Big\}\,,
\nonumber
\ea
where
\ba
J_{\cal A,B,C}(x,A,k) &=&
\int_{-A}^{A} \frac{dy}{\sqrt{1+y}}\, e^{-x\sqrt{1+y}}\, P_{\cal A,B,C}(y/A,k)
\\
I_{\cal D}(x,A,k) &=& \int_{-A}^A dy \,e^{-x\sqrt{1+y}}\,  P_{\cal D}(y/A,k).
\ea
Via the change of variables $u=\sqrt{1+y}$,
these integrals can be obtained in terms of the function
\be
g(x,A) \equiv \frac{1}{x}(e^{-x\sqrt{1-A}}-e^{-x\sqrt{1+A}}).
\ee
We have 
\ba
J_{\cal A,B,C}(x,A,k)&=&  2 
P_{\cal A,B,C}\Big(A^{-1}({\txts\frac{d^2}{dx^2}}-1),k\Big)\, g(x,A) 
\\
I_{\cal D}(x,A,k)&=& -2 \frac{d}{dx} \,
P_{\cal D}\Big(A^{-1}({\txts\frac{d^2}{dx^2}}-1),k\Big)\, g(x,A) 
\ea
Note that $\sqrt{1\pm A} = \frac{|k\pm 1|}{\sqrt{k^2+1}}$
for $A=\frac{2k}{k^2+1}$.

One then obtains the correlator in the form
\be
\int d^3x\,e^{i{\bf q\cdot x}}\<T_{12}(0)\,T_{12}(x)\> 
= \frac{q^5}{\pi^2} \sum_{n,m\in{\bf Z}} 
{\rm g}
\left([\Pi];~s_m({\txts\frac{x_0}{L_0}})\,s_n({\txts\frac{x_0}{L_0}}),~q|x_0+mL_0|,~q|x_0+nL_0|\right),
\la{eq:CE=sum}
\ee
where 
\be
 {\rm g}([\Pi];s,x,y) = \frac{e^{-x}\,\Pi(s,x,y) - e^{-y}\,\Pi(s,y,x)}{(x^2-y^2)^5} = 
{\rm g}([\Pi];s,y,x)
\ee
and the appropriate polynomials $\Pi$  will be specified below 
for a few cases of interest. The function ${\rm g}$ 
will always be finite at $x=y$, and we give the first two terms in
the expansion around this point, which is useful when
implementing expression (\ref{eq:CE=sum}) numerically.
Since the series in $n$ and $m$ are exponentially convergent,
only a few terms are needed to reach, say, a precision of $10^{-6}$.

Consider the case ${\bf q}=(q,0,0)$, $q\geq 0$.
Relabelling the p-coordinates $p_1\to p_3$, $p_3\to p_2$,
$p_2\to p_1$ in \eq\ref{eq:poisson} before going to spherical 
coordinates, we obtain, 
\ba
P_{\cal A}(y,k) &=& (1-y^2)\,\Big(y^2+\frac{4}{k}y+\frac{2}{k^2}+1\Big)\,,
\\
P_{\cal B}(y,k) &=&  y \,(1-y^2)\,\Big(y+\frac{1}{k}\Big)\,,
\\
P_{\cal C}(y,k) &=& 1-y^4\,,
\\
P_{\cal D}(y,k) &=& 1-y^2\,,
\ea
and from there
\ba
2 \,  \Pi_{12,12}^{\hat e_1}(s,x,y) &=& 4608
    +  x(6+x)(768+x(240+x(72+x(10+x))))
\\
&&   -4sx(48+x(48+x(12+x)))y
\nonumber
\\
&&
   -(96+x(96+x(120+x(32+3x))))y^2
\nonumber
\\
&&   +8sx(6+x)y^3
   +(6+x)(-2+3x)y^4
   -4sy^5
   -y^6\,,
\nonumber
\ea
\ba
&&{\rm g}([\Pi_{12,12}^{\hat e_1}];s,x+\half\epsilon,x-\half\epsilon) =
\frac{e^{-x}}{240x^5} 
\Big\{ x^3(2+s)+3(7+5s)+3x(7+5s)+x^2(9+6s)
\nonumber
\\
&& \qquad \frac{\epsilon^2}{56} \Big((39+21s+x^3(4+s)+3x(13+7s)+x^2(17+8s))   \Big) 
+{\rm O}(\epsilon^4)\Big\}\,.
\la{eq:g_eq}
\ea
In \eq\ref{eq:g_eq}, we have defined ${\rm g}$ by continuity.

 For the case ${\bf q}=(0,0,q)$, $q\geq 0$, we obtain
\ba
P_{\cal A}(y,k) &=& \frac{1}{4}y^4 + \frac{2}{k}y^3 + \Big(\frac{1}{k^2}
           +\frac{3}{2}\Big)y^2 + \frac{2}{k}y + \frac{1}{k^2}+\frac{1}{4}\,,
\\
P_{\cal B}(y,k) &=& \frac{1}{4}(1-y^2)^2\,,
\\
P_{\cal C}(y,k) &=& P_{\cal A}(y)\,,
\\
P_{\cal D}(y,k) &=&  2y\,(y+\frac{1}{k})\,,
\ea
and further
\ba
\Pi_{12,12}^{\hat e_3}(s,x,y) &=& -576 
             - x(4+x)(144+x(48+x(24+x(6+x)))) 
\\
&& + 2sx(96+x(96+x(48+x(11+x))))y
\nonumber
\\
&& + (-48+x(-48+x(-48+(-4+x)x)))y^2
\nonumber
\\
&& -4sx^2(5+x)y^3
   + x(14+x)y^4
    + 2s(-1+x)y^5
     -y^6
\,.
\nonumber
\ea
\ba
&& {\rm g}([\Pi_{12,12}^{\hat e_3}];s,x,x) = \frac{e^{-x}}{240x^5}
\,\Big\{ 21+x(21+x(9+x(2+x)))
\\
&& \qquad +s(15-x(-15+(-1+x)x(3+x)))
\nonumber
\\
&&\qquad  \frac{\epsilon^2}{56} 
\Big( 69-2x^3(-1+s)-x^4(-1+s)+63s
\nonumber
\\
&&\qquad  +x^2(25+19s)+x(69+63s) \Big) +{\rm O}(\epsilon^4)\Big\}\,.
\nonumber
\ea
Note that the polynomials $P_{\cal A,B,C,D}(y,k)$ all have 
the symmetry $P_{\cal A,B,C,D}(-y,-k)=P_{\cal A,B,C,D}(y,k)$.

\subsection{$\<T_{11}T_{11}\>$ with momentum along $x$ direction}
For $\int d^3{\bf x} e^{i{\bf q\cdot x}} \<T_{11}(0)T_{11}(x)\>$,
the polynomial is
\ba
-\Pi_{11,11}^{\hat e_1}(s,x,y) &=&
 4608
+   x(4608+x(2112+x(576+x(108+x(14+x)))))
\nonumber
\\
&& +  sx(2+x)(120+x(60+x(12+x))) y
\nonumber
\\
&&   -(192+x(192+x(24+x(4+x))))y^2
   -2sx(18+x(8+x))y^3
\nonumber
\\
&&   -(-12+x(10+x))y^4
+   s(2+x)y^5
+   y^6\,,
\ea
\ba
&&{\rm g}([\Pi_{11,11}^{\hat e_1}];s,x+\half\epsilon,x-\half\epsilon)
 = \frac{e^{-x}}{480x^5}
\\
&&\qquad \Big\{ 66(1+x)+2x^2(15+x(4+x))+  s(30+x(30+x(15+x(5+x))))
\nonumber
\\
&&\qquad
+\frac{\epsilon^2}{56}\Big(138+138x +x^4(2+s)+x^3(8+5s)+x^2(54+5s)\Big)
+{\rm O}(\epsilon^4)\Big\}\,.
\nonumber
\ea

\subsection{Scalar correlator with non-zero momentum}
For $\frac{1}{4}\sum_{\mu<\nu,\rho<\sigma} \int d^3{\bf x}\,e^{i{\bf q\cdot x}} 
\<F_{\mu\nu}^2(0)F_{\rho\sigma}^2(x)\>$, the polynomial is 
\ba
-\Pi_{\theta,\theta}^{\hat e_3}(s,x,y) &=&
x^4(12+x(6+x))
 -2sx^3(24+x(9+x))y
\\
&& -x^2(-72+(-12+x)x)y^2
 +4sx(-12+x(3+x))y^3
\nonumber
\\
&&  -(-12+x(18+x))y^4
 -2s(-3+x)y^5
+y^6.
\nonumber
\ea
\ba
&&{\rm g}([\Pi_{\theta,\theta}^{\hat e_3}];s,x+\half\epsilon,x-\half\epsilon) =
 \frac{e^{-x}}{240 x^5}  \Big\{ 
-x^4(-1+s)+45(1+s)+45x(1+s)+15x^2(1+s)
\nonumber
\\
&&\qquad
+\frac{\epsilon^2}{56}
\Big( -x^4(-1+s)+105(1+s)+105x(1+s)+35x^2(1+s) \Big)
+{\rm O}(\epsilon^4)  \Big\}\,.
\ea

\section{Calculation of spectral functions}
In this appendix, 
we show how to obtain \eq(\ref{eq:sfqfirst}--\ref{eq:sfqlast})
in the case of the $\<T_{12}T_{12}\>$ correlator.
Other cases can be treated in exactly the same way.
In this section $q$ is understood to be in units of $2T$.
Using the propagator in the mixed representation,
\be
\frac{1}{L_0}\sum_{p_0} \frac{e^{ip_0x_0}}{p_0^2+{\bf p}^2} =  \frac{1}{2|{\bf p}|}
\frac{\cosh |{\bf p}|(\half L_0-x_0)}{\sinh \half |{\bf p}| L_0},
\la{eq:prop}
\ee
we obtain ($\tau=1-2Tx_0$)
\ba
{\cal A,B,C} &=& \frac{1}{\pi^2L_0^5}
\int_0^\infty p^4dp\int_{-1}^1 dx 
\frac{p}{p^2+ q^2+2p qx}\, P_{\cal A,B,C}(x,{\txts\frac{p}{ q}})
\\ 
&& \frac{\cosh(p+\sqrt{p^2+ q^2+2p qx})\tau + \cosh(p-\sqrt{p^2+ q^2+2p qx})\tau}
     {\cosh(p+\sqrt{p^2+ q^2+2p qx}) - \cosh(p-\sqrt{p^2+ q^2+2p qx}) }.
\nonumber
\ea
For the term proportional to $\cosh(p+\sqrt{p^2+q^2+2pqx})\tau$,
we do the change of variables $p=\frac{p^2-q^2}{2(k+qx)}$,
which brings this expression to the desired form $\cosh(k\tau)$
on the integration interval $k|_q^\infty$.

For the $\cosh(p-\sqrt{p^2+q^2+2pqx})\tau$ term,
we choose $p=\frac{k^2-q^2}{2(qx-k)}$, which achieves 
the same on the integration interval $k|_{qx}^q$.
We then notice that if ${\cal F}(-x,-k,q)={\cal F}(x,k,q)$,
\[ 
\int_{-1}^1  dx \int_{qx}^q dk {\cal F}(x,k,q) = \int_0^q \int_{-1}^1
{\cal F}(x,-k,q).
\]
One therefore finds that the first term determines the 
spectral function above the threshold $q$, while the second 
determines it below $q$. We write 
\be
\rho(\omega,q,T) = \rho_<(\omega,q,T) + \rho_>(\omega,q,T),
\ee
where the first term vanishes for $\omega>q$ and the second 
for $\omega<q$.

After some algebra, 
\ba
\rho_{{\cal A,B,C},>}(2kT,2qT,T)& =&
\frac{T^4\theta(k-q)\sinh k (k^2-q^2)^5}{64\pi^2}
\\ 
&& \int_{-1}^1 dx\frac{P_{\cal A,B,C}(x,\frac{k^2-q^2}{2q(k+qx)})}{(k+qx)^6}
\frac{1}{\cosh k -\cosh q\frac{q+kx}{k+qx}}.
\nonumber
\ea
and after the further change of variables $x=\frac{kz-q}{k-qz}$, 
we obtain
\ba
\rho_{{\cal A,B,C},>}(2kT,2qT,T)&=&\frac{T^4\theta(k-q)\sinh k}{64\pi^2}
\\ 
&& \int_{-1}^1 \frac{dz(k-qz)^4}{\cosh k-\cosh qz}\,
P_{\cal A,B,C}\Big(\frac{kz-q}{k-qz},\frac{k-qz}{2q}\Big).
\nonumber
\ea

After the change of variables $p=p(k)$ but
before the change of variables $x=x(z)$, the sub-threshold part,
$\rho_<(2kT,2qT,T)$, has the same expression as $\rho_>$.
However, in order to perform the change of variables $x(z)$,
the integral must first be split, $\int_{-1}^1dx=\int_{-1}^{-k/q} dx+\int_{-k/q}^1 dx$.
These integration intervals are mapped to the $z$-integrals $\int_{-\infty}^{-1}$ and 
$\int_1^{\infty}$. Because $\frac{dx}{dz}<0$ on these two intervals, 
a minus sign relative to $\rho_>$ appears. One thus finds
\ba
\rho_{{\cal A,B,C},<}(2kT,2qT,T)&=& \frac{T^4\theta(q-k)\sinh k}{64\pi^2}
\left( \int_{-\infty}^{-1}+\int_1^\infty  \right) 
\\ 
&& \frac{dz(k-qz)^4}{\cosh qz-\cosh k}\,
P_{\cal A,B,C}\Big(\frac{kz-q}{k-qz},\frac{k-qz}{2q}\Big).
\nonumber
\ea

For the contribution 
\be
{\cal D} = \frac{T^5}{\pi^2} \int_{-1}^1 \int_0^\infty dp p^4
P_{\cal D}(x,{\txts\frac{p}{q}}) 
\frac{\cosh(p+\sqrt{p^2+ q^2+2p qx})\tau - \cosh(p-\sqrt{p^2+ q^2+2p qx})\tau}
     {\cosh(p+\sqrt{p^2+ q^2+2p qx}) - \cosh(p-\sqrt{p^2+ q^2+2p qx}) }
\la{eq:Drho1}
\ee
we follow exactly the same steps 
as for ${\cal A,B,C}$ and find
\ba
\rho_{{\cal D},>}(2kT,2qT,T)&=&
\frac{T^4\theta(k-q)\sinh k}{64\pi^2}
\\  
&& \int_{-1}^1 \frac{dz(k-qz)^2(k^2-q^2z^2)}{\cosh k-\cosh qz}\,
P_{\cal D}\Big(\frac{kz-q}{k-qz},\frac{k-qz}{2q}\Big).
\nonumber
\ea
For the sub-threshold part  (which includes
the minus sign in the numerator of \eq\ref{eq:Drho1}) we find 
the same expression as for $\rho_>$ before the change of variables $x(z)$.
Therefore, again, in the final expression 
$\rho_<$ has a relative minus sign as well as the complementary
integration range as compared to $\rho_>$,
\ba
\rho_{{\cal D},<}(2kT,2qT,T)&=&
 \frac{T^4\theta(q-k)\sinh k}{64\pi^2}
\left(\int_{-\infty}^{-1}+\int_1^\infty\right) 
\\ 
&& \frac{dz(k-qz)^2(k^2-q^2z^2)}{\cosh qz-\cosh k}\,
P_{\cal D}\Big(\frac{kz-q}{k-qz},\frac{k-qz}{2q}\Big).
\nonumber
\ea
Finally, in view of the symmetry $P(-y,-k)=P(y,k)$ of the relevant
polynomials, $\int_{-1}^1$ can be replaced by $2\int_0^1$ and 
$\int_{-\infty}^{-1}+\int_{1}^\infty$ by $2\int_{1}^\infty$.

\section{The spectral function in terms of polylogarithms\la{app:SF}}
Polylogarithms are defined by the series ${\rm Li}_n(z)=\sum_{k\geq1}\frac{z^k}{k^n}$
and analytic continuation thereof.
The integrals appearing in the spectral function are given as follows
in terms of these special functions:
\ba
\la{eq:Iz2}
{\rm q}^3\,{\cal I}([z^2];2T{\rm w}, 2T{\rm q}, T)&=&
-{\rm q}^2\left((\log(1-e^{{\rm q}-{\rm w}})-\log(e^{{\rm w}+{\rm q}}-1) \right)                   \\ && 
-2{\rm q}\,\left({\rm Li}_2(e^{{\rm q}-{\rm w}}) -{\rm Li}_2(e^{{\rm w}+{\rm q}}) \right) \,        
+2\,\left({\rm Li}_3(e^{{\rm q}-{\rm w}}) -{\rm Li}_3(e^{{\rm w}+{\rm q}})\right)         \nn &&
+i\pi {\rm q}^2 -\frac{1}{3}\,\left(1+\theta({\rm q}-{\rm w})\right)
\left( {\rm w}^3-2\pi^2 {\rm w} + 3i\pi {\rm w}^2 \right)               \nonumber
\\
\la{eq:Iz4}
{\rm q}^5\,{\cal I}([z^4];2T{\rm w}, 2T{\rm q}, T)&=&
-{\rm q}^4\left(\log(1-e^{{\rm q}-{\rm w}})-\log(e^{{\rm q}+{\rm w}}-1)\right)           \\ && 
-4{\rm q}^3 \left({\rm Li}_2(e^{{\rm q}-{\rm w}})-{\rm Li}_2(e^{{\rm w}+{\rm q}})\right)    \,
+12{\rm q}^2\left({\rm Li}_3(e^{{\rm q}-{\rm w}})-{\rm Li}_3(e^{{\rm q}+{\rm w}})\right)    \nn &&
-24{\rm q} \left({\rm Li}_4(e^{{\rm q}-{\rm w}})-{\rm Li}_4(e^{{\rm q}+{\rm w}})\right)      \,
+24 \left( {\rm Li}_5(e^{{\rm q}-{\rm w}})-{\rm Li}_5(e^{{\rm q}+{\rm w}})\right)     \nn &&
+ i\pi {\rm q}^4 -\frac{1}{15}\left(1+\theta({\rm q}-{\rm w})\right)
\left( 3{\rm w}^5 - 20\pi^2{\rm w}^3 - 8\pi^4{\rm w} + 15i\pi {\rm w}^4  \right)  \nonumber
\ea
The integral with $z^0$ in the numerator can be expressed in 
terms of elementary functions, see \eq\ref{eq:sfelem}.
Hence in all cases the expressions on the left and on the right
of the threshold differ by a polynomial in $\omega$.
In order to simplify expressions \eq(\ref{eq:Iz2}) and (\ref{eq:Iz4}), we have used the identities
\ba
{\rm Li}_5(e^{-{\rm w}})-{\rm Li}_5(e^{\rm w})  &=& 
\frac{{\rm w}^5}{120} - \frac{\pi^2}{18}{\rm w}^3-\frac{\pi^4}{45}{\rm w}+\frac{i\pi}{24}{\rm w}^4
\\
{\rm Li}_3(e^{-{\rm w}})-{\rm Li}_3(e^{\rm w})  &=&
\frac{{\rm w}^3}{6}-\frac{\pi^2}{3}{\rm w}+
\frac{i\pi}{2}{\rm w}^2\,.
\ea

\bibliographystyle{JHEP}
\bibliography{viscobib}
\end{document}